\documentclass[fleqn,usenatbib]{mnras}
\usepackage{newtxtext,newtxmath}

\usepackage[T1]{fontenc}
\usepackage{ae,aecompl}

\usepackage{graphicx}	
\usepackage{amsmath}	
\usepackage{xspace}        
\usepackage[para,online,flushleft]{threeparttable}

\newcommand{\target}{NGC 4395}
\newcommand{\mjyb}{mJy~beam$^{-1}$}

\title[Radio nucleus in NGC 4395]{Is there a sub-parsec-scale jet base in the nearby dwarf galaxy NGC 4395?}

\author[J. Yang et al.]{Jun Yang$^{1}$\thanks{E-mail: jun.yang@chalmers.se}, 
Xiaolong Yang$^{2}$\thanks{E-mail: yangxl@shao.ac.cn},
Joan M. Wrobel$^{3}$,
Zsolt Paragi$^{4}$,
Leonid I. Gurvits$^{4,5}$,
Luis C. Ho$^{6,7}$,
\and
Kristina Nyland$^{8}$,
Lulu Fan$^{9,10}$
and Daniel Tafoya$^{1}$
\\
$^{1}$Department of Space, Earth and Environment, Chalmers University of Technology, Onsala Space Observatory, SE-439 92 Onsala, Sweden \\
$^{2}$Shanghai Astronomical Observatory, Key Laboratory of Radio Astronomy, Chinese Academy of Sciences, 200030 Shanghai, China \\
$^{3}$National Radio Astronomy Observatory, P.O. Box O, Socorro, NM 87801, USA \\
$^{4}$Joint Institute for VLBI ERIC (JIVE), Oude Hoogeveensedijk 4, 7991 PD Dwingeloo, The Netherlands \\
$^{5}$Faculty of Aerospace Engineering, Delft University of Technology, Kluyverweg 1, 2629 HS Delft, The Netherlands \\
$^{6}$Kavli Institute for Astronomy and Astrophysics, Peking University, 100871 Beijing, China \\
$^{7}$Department of Astronomy, School of Physics, Peking University, 100871 Beijing, China \\
$^{8}$U.S. Naval Research Laboratory, 4555 Overlook Ave SW, Washington, DC 20375, USA \\
$^{9}$CAS Key Laboratory for Research in Galaxies and Cosmology, Department of Astronomy, University of Science and Technology of China, 230026 Hefei, China \\
$^{10}$School of Astronomy and Space Sciences, University of Science and Technology of China, 230026 Hefei Anhui, China 
}
\date{Accepted XXX. Received YYY; in original form ZZZ}

\pubyear{2021}

\begin{document}
\label{firstpage}
\pagerange{\pageref{firstpage}--\pageref{lastpage}}
\maketitle
\begin{abstract}
NGC~4395 is a dwarf galaxy at a distance of about 4.3~Mpc (scale: $\sim$0.021 pc~mas$^{-1}$). It hosts an intermediate-mass black hole (IMBH) with a mass between $\sim$10$^4$ and $\sim$10$^5$ solar masses. The early radio observations of NGC~4395 with the very long baseline interferometry (VLBI) network, High Sensitivity Array (HSA), at 1.4~GHz in 2005 showed that its nucleus has a sub-mJy outflow-like feature (E) extending over 15~mas. To probe the possibility of the feature E as a continuous jet with a base physically coupled with the accretion disc, we performed deep VLBI observations with the European VLBI Network (EVN) at 5~GHz, and analysed the archival data obtained with the HSA at 1.4~GHz in 2008, NSF's Karl G. Jansky Very Large Array (VLA) at 12--18~GHz and the Atacama Large Millimetre/submillimetre Array (ALMA) at 237~GHz. The feature E displays more diffuse structure in the HSA image of 2008 and has no compact substructure detected in the EVN image. Together with the optically thin steep spectrum and the extremely large angular offset (about 220~mas) from the accurate optical \textit{Gaia} position, we explain the feature E as nuclear shocks likely formed by the IMBH's episodic ejection or wide-angle outflow. The VLA and ALMA observations find a sub-mJy pc-scale diffuse feature, possibly tracing a thermal free-free emission region near the IMBH. There is no detection of a jet base at the IMBH position in the VLBI maps. The non-detections give an extremely low luminosity of $\leq4.7\times10^{33}$~erg\,s$^{-1}$ at 5~GHz and indicate no evidence of a disc-jet coupling on sub-pc scales.

\end{abstract}




\section{Introduction}
\label{sec1}

Low-mass galaxies including dwarf galaxies (stellar mass $M_\mathrm{\star}\leq10^{9.5}M_{\sun}$) are expected to host black holes (BHs) with masses close to their ``birth'' values \citep[e.g.][]{Greene2020, Volonteri2021} because they underwent less merging and intensive accretion events than their larger counterparts. These low-mass BHs with masses $10^2 M_{\sun} \leq M_\mathrm{bh} \leq 10^6 M_{\sun}$ are usually classified as intermediate-mass black holes (IMBHs). Hunting for IMBHs in low-mass galaxies can help to probe the co-evolution of galaxies and massive BHs \citep[e.g.][]{Greene2006, Greene2007, Kormendy2013, Baldassare2020, Reines2022}. 

When IMBHs accrete, they can reveal themselves as low-mass active galactic nuclei (AGNs). Currently, there are several hundred IMBH candidates (fraction $<1$ per cent of low-mass galaxies) found by optical and X-ray observations of low-mass galaxies \citep[e.g.][]{Greene2007, Reines2013, Pardo2016}. However, only a few of these IMBH candidates \citep{Greene2020} have sufficiently robust mass constraints for them to be considered as bona fide IMBHs \citep[e.g.][]{Greene2006, Baldassare2015, Woo2019}.

\target{} hosts one of the lowest-mass accreting IMBHs known. \target{} is a nearby well-resolved dwarf galaxy at redshift $z=0.00106$ and hosts an extremely low-luminosity Seyfert 1 nucleus \citep[][]{Filippenko1989, Filippenko1993, Ho1995}. \citet{Filippenko2003} first pointed out that the AGN might be energized by an IMBH with a mass between $\sim 10^4 M_{\sun}$ and $\sim 10^5 M_{\sun}$. Estimates of the IMBH mass are presented in several follow-up studies  \citep[e.g.][]{Peterson2005, Edri2012, DenBrok2015, LaFranca2015, Brum2019, Woo2019}. However these mass estimates still have relatively large uncertainties. The early optical reverberation observations found that the central IMBH in \target{} has a mass of $M_{\rm bh} = (3.6\pm1.1)\times10^5 M_{\sun}$ \citep{Peterson2005}. Recently, \citet{Woo2019} have performed another reverberation-based mass measurement and reported a smaller mass, $M_{\rm bh} = 10^{4.0\pm0.4} M_{\sun}$ (including systematic errors).

Jets are frequently found in supermassive and stellar-mass BH accretion systems \citep[cf. a review by][]{Blandford2019}. The innermost parts of continuous radio jets are referred to as jet bases, sometimes also called radio cores. They are partially optically thick and have relatively flat spectra at frequencies $\la$10~GHz. Detections of jet bases provide direct support for a physical coupling between the jets and their BH accretion discs \citep[e.g.][]{Fender2004, Merloni2003, Fischer2021}. 

IMBHs are also expected to launch continuous radio jets. About 0.3 per cent of dwarf galaxies from the sample of \citet{Reines2020} have radio counterparts in the FIRST \citep[Faint Images of the Radio Sky at Twenty Centimetres,][]{Becker1995} survey. High-resolution observations, including using the very long baseline interferometry (VLBI) technique, have been performed to search for jets that provide independent evidence of IMBH accretion and ejection activity in a sample of dwarf galaxies with wandering IMBH candidates \citep{Sargent2022} and some nearby low-mass galaxies, e.g. GH~10 \citep{Greene2006, Greene2006RQ, Wrobel2008}, POX~52 \citep{Thornton2008}, ESO~243$-$49 HLX-1 \citep{Web2012, Cseh2015}, Henize~2--10 \citep{Reines2012, Schutte2022}, Mrk~709 \citep{Reines2014}, and NGC~404 \citep[][]{Paragi2014, Nyland2017, Davis2020}. VLBI detections of compact radio cores would provide data points for filling the mass gap between supermassive and stellar-mass BHs \citep[e.g.][]{Greene2020} and probing some mass-dependent scaling relations \citep[e.g. the fundamental plane relation, ][]{Merloni2003, Saikia2018, Fischer2021}. To date, pc-scale compact radio components have been reported in a few dwarf galaxies: the candidate satellite galaxy of NGC~5252 \citep{Yang2017, Mezcua2018, Kim2020}, SDSS J090613.77$+$561015.2 \citep{Yang2020IMBH} and NGC~4395 \citep{Wrobel2001, Wrobel2006}. However, there is no evidence of a continuous pc-scale jet with a flat-spectrum radio core in dwarf AGNs. These observational data on the IMBH jets play a key role as starting points for future studies of the IMBH population at radio wavelengths with the next-generation arrays \citep[e.g.][]{Greene2020, Liodakis2022}.

\begin{figure}
\centering
\includegraphics[width=\columnwidth]{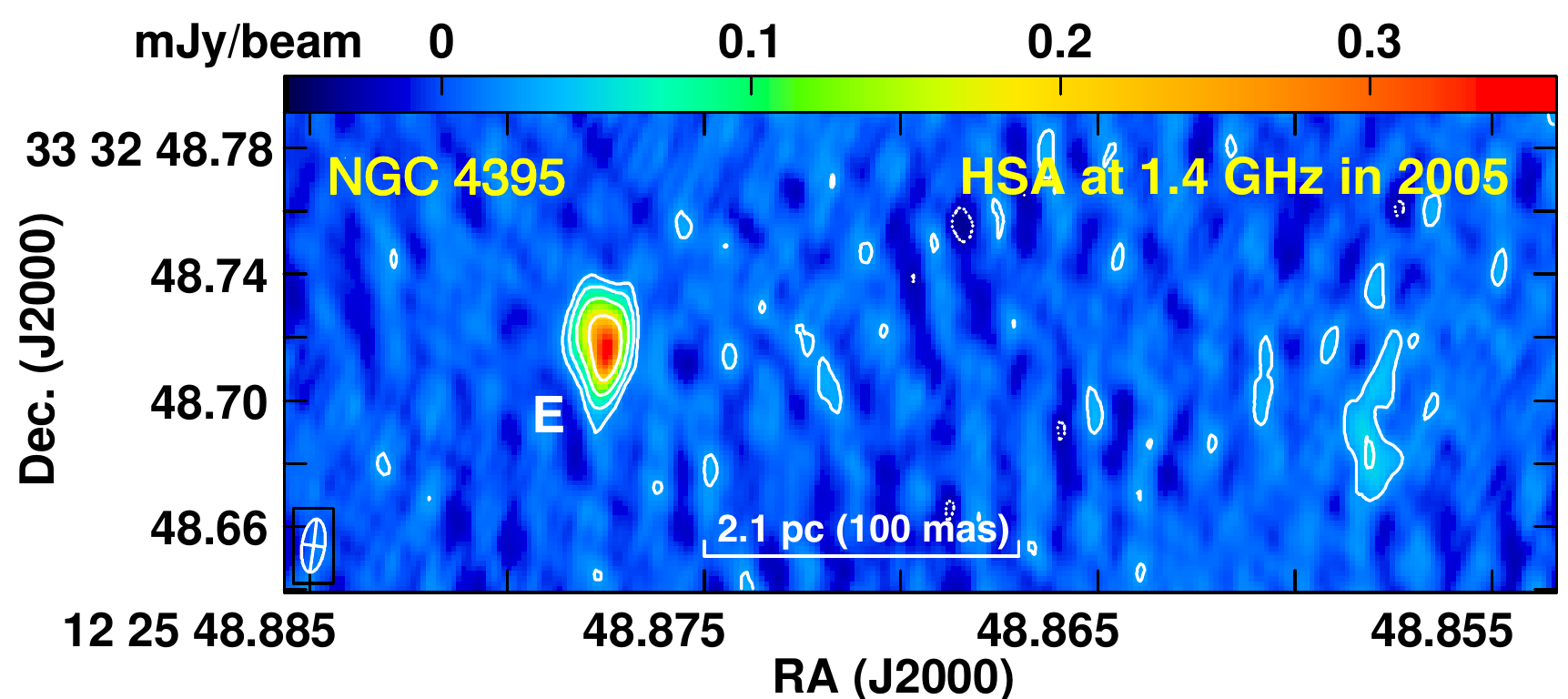}  \\
\caption{
The sub-pc-scale outflow-like feature of \target{} observed previously by \citet{Wrobel2006} with the HSA at 1.4 GHz on 2005 May 1. The contours start from 0.025~mJy\,beam$^{-1}$ (2.5$\sigma$) and increase by factors $-$2, $-$1, 1, 2, 4, 8. The peak brightness is 0.357~mJy\,beam$^{-1}$. The full width at half maximum (FWHM) of the synthesised beam is $17.0 \times 7.2$~mas at position angle $-8\fdg0$. }
\label{fig:hsa_2005}
\end{figure}

\target{} has been observed by various radio arrays: the historical Very Large Array (VLA) at multiple frequencies \citep[e.g.][]{Sramek1992, Moran1999, Becker1995, Ho2001}, NSF's Karl G. Jansky VLA with very wide bandwidths \citep[$\ge$2~GHz, ][]{King2013, Saikia2018, Lacy2020}, the enhanced Multi-Element Remotely Linked Interferometre Network (e-MERLIN) at 1.5~GHz \citep{Baldi2021}, the Very Long Baseline Array (VLBA) at 1.4~GHz \citep{Wrobel2001} and the High Sensitivity Array (HSA) at 1.4~GHz \citep{Wrobel2006}. Since the host galaxy has some diffuse radio emission, the observed total flux density shows significant dependence on the image resolution. At 1.4~GHz, its radio nucleus shows a compact structure with a flux density of 1.17$\pm$0.15~mJy in the FIRST image \citep{Becker1995}. In the high-resolution 1.4-GHz HSA image, it displays an outflow-like feature extending over 15~mas with a total flux density of 0.74$\pm$0.05~mJy \citep{Wrobel2006}. That HSA image from 2005 is also shown in Fig.~\ref{fig:hsa_2005}. The simultaneous radio and X-ray monitoring observations of \target{} \citep{King2013} displayed a hint for a pc-scale disc–jet coupling in the low-mass AGN and implied the existence of a jet base. To investigate whether the elongated feature includes a jet base powered by the central accreting IMBH, we conducted VLBI observations with the EVN at 5.0~GHz and re-analysed the existing archival data observed with the HSA at 1.4~GHz in 2008, the Jansky VLA at 12--18~GHz in 2016 and the Atacama Large Millimetre/submillimetre Array (ALMA) at 237.1~GHz in 2018--2019. 

This paper is organised as follows. We introduce the radio observations and the data reduction in Section~\ref{sec:observations}, present multi-frequency imaging results of \target{} in Section~\ref{sec:results}, and discuss radio and optical astrometry precisions, describe the interpretation of the multi-scale radio morphology and present potential implications in Section~\ref{sec:discussion}. We give our conclusions in Section~\ref{sec:conclusion}. Throughout the paper, a standard $\Lambda$CDM cosmological model with H$_\mathrm{0}$~=~71~km~s$^{-1}$~Mpc$^{-1}$, $\Omega_\mathrm{m}$~=~0.27, $\Omega_{\Lambda}$~=~0.73 is adopted. For our target, this gives a luminosity distance of 4.5~Mpc, fully in agreement with the Cepheid distance $4.3\pm0.3$~Mpc presented by \citet{Thim2004}. We define the spectral index $\alpha$ with the power-law spectrum $S(\nu) \propto \nu^{\alpha}$.

\section{Observations and data reduction}
\label{sec:observations}
In this Section, the related radio observations and the methods of our data reduction are reported. We give the details on the high-resolution VLBI observations and the data reduction in Subsections~\ref{subsec:EVNobs} and \ref{subsec:HSAobs}, summarise the VLBI experiment setups in Table~\ref{tab:obs} and list some basic information on the VLBI calibrators in Table~\ref{tab:pos4cal}. Furthermore, we describe the high-frequency VLA and ALMA observations and data calibration in Subsections~\ref{subsec:VLAobs} and \ref{subsec:ALMAobs}, and list the setup and the calibrator information of the multi-epoch ALMA observations in Table~\ref{tab:ALMAobs}.      

\subsection{The EVN experiment at 5 GHz}
\label{subsec:EVNobs}
We observed \target{} with all the available EVN and \textit{e}-MERLIN telescopes at 5.0~GHz for 6~h during the \textit{e}-EVN session 2022 January 18--19. There were 17 participating telescopes: Jodrell Bank Lovell (\texttt{JB1}) and Mk2 (\texttt{JB2}), Westerbork (\texttt{WB}, single dish), Effelsberg (\texttt{EF}), Medicina (\texttt{MC}), Tianma (\texttt{T6}), Onsala (\texttt{ON}), Toru\'n (\texttt{TR}), Yebes (\texttt{YS}), Svetloe (\texttt{SV}), Zelenchukskaya (\texttt{ZC}), Badary (\texttt{BD}), Hartebeesthoek (\texttt{HH}), Cambridge (\texttt{CM}) Knockin (\texttt{KN}), Pickmere (\texttt{PI}) and Defford (\texttt{DE}). All the EVN stations used dual circular polarisation and 2-bit quantisation. The observations were performed with a maximum date rate of 2048~Mbps (16 32-MHz sub-bands). Due to the limited network bandwidths, the Russian stations (\texttt{SV}, \texttt{Zc} \& \texttt{BD}) had 8 32-MHz sub-bands and the \textit{e}-MERLIN stations (\texttt{JB1}, \texttt{CM}, \texttt{KN}, \texttt{PI} \& \texttt{DE}) gave 2 64-MHz sub-bands. The data correlation was done in real time by the EVN software correlator SFXC \citep[][]{Keimpema2015} at JIVE (Joint Institute for VLBI ERIC) using the typical correlation parameters for continuum experiments: 0.5-MHz frequency resolution and 1-s integration time.

\begin{table*}
\caption{Summary of the EVN and HSA observations of \target{}.  }
\label{tab:obs}
\begin{tabular}{cccccc}
\hline
Project & Starting time (duration)     & Freq.     & Rate      & Participating VLBI stations       \\     
code    &   (UTC)                      & (GHz)     & (Mbps)    &  (see Sect. \ref{sec:observations} for the definition of the station codes)   \\
\hline     
EY039   & 2022 Jan 18, 23.0~h (6~h)    & 4.99      & 2048      & \texttt{JB1, JB2, WB, EF, MC, T6, ON, TR, YS, SV, ZC, BD, HH, CM, KN, PI, DE}  \\    
BW089   & 2008 May 04, 23.5~h (8~h)    & 1.41      & 512       & \texttt{BR, EF, FD, GB, HN, KP, LA, MK, OV, PT, SC, Y27, AR}    \\
\hline
\end{tabular}
\end{table*}

\begin{table*}
\caption{Summary of the VLBI phase-referencing calibrators. The positional error $\sigma_{\rm pos}$ is estimated with respect to the ICRF3 \citep{Charlot2020}. }
\label{tab:pos4cal}
\begin{tabular}{ccccccccc}
\hline
Project  & Array & Calibrator            & RA                                   & Dec.                                      & $\sigma_{\rm pos}$ 
                                                                                                                                     & $S_{\rm int}$ & Separation & Structure \\     
code     &       &                       &   (J2000)                            & (J2000)                                   & (mas)  & (mJy)         &  ($\degr$) & \\
\hline     
EY039    & EVN   & J1220$+$3343          & 12$^{\rm h}$20$^{\rm m}$33$\fs$87550 & $+$33$\degr$43$\arcmin$12$\farcs$0278     & 0.2    & $125\pm6$     &  1.1       & One-sided core-jet source  \\
EY039    & EVN   & J122755.50$+$334526.9 & 12$^{\rm h}$27$^{\rm m}$55$\fs$48611 & $+$33$\degr$45$\arcmin$26$\farcs$9515     & 0.2    &  $20\pm1$     &  0.5       & Slightly elongated jet \\
BW089    & HSA   & J1220$+$3431          & 12$^{\rm h}$20$^{\rm m}$08$\fs$29416 & $+$34$\degr$31$\arcmin$21$\farcs$7427     & 0.8    & 190$\pm$9     &  1.5       & One-sided core-jet source  \\
\hline
\end{tabular} \\

\end{table*}

The VLBI observations were conducted in the phase-referencing mode \citep[e.g.][]{Rioja2020}. We used J1220$+$3343 as the main phase-referencing calibrator. This source is one of those that define the third realisation of the International Celestial Reference Frame \citep[ICRF3, ][]{Charlot2020}. Its position is listed in Table~\ref{tab:pos4cal}. The calibrator is separated by $1\fdg1$ from the target, has a flat radio spectrum and shows a core-jet structure with a correlated amplitude of $\geq$40~mJy on the long baselines in the geodetic VLBI observations at 2.3 and 8.4~GHz \citep{Charlot2020}. We observed the pair of sources with a cycle time of 8~min (50~s for the calibrator, 390~s for the target, 40~s for two slewing gaps). To improve the phase-referencing calibration further, we also observed a faint source VLASS J122755.50$+$334526.9 \citep{Gordon2021} as a nearby calibrator for one 2.5-min scan per half hour. The faint calibrator is separated by 29~arcmin from \target{}, and has a rising spectrum with total flux densities of $\sim$17~mJy in the Karl G. Jansky Very Large Array Sky Survey (VLASS) at 2--4~GHz \citep{Lacy2020, Gordon2021} and $\sim$12~mJy in the historical VLA survey FIRST \citep{Becker1995}.

The National Radio Astronomy Observatory (NRAO) Astronomical Image Processing System \citep[\textsc{aips} version 31DEC21,][]{Greisen2003} software package was used to calibrate the visibility data. We followed the general EVN data calibration strategy. First, the no-fringe data were flagged out. Second, a priori amplitude calibrations were performed. Third, the ionospheric dispersive delays were removed. Fourth, the phase errors due to the antenna parallactic angle variations were corrected. Fifth, the instrumental phase and delay offsets across sub-bands were solved via manual fringe fitting on the 2-min data of the source OQ~208. Sixth, the fringe-fitting on the calibrator data was performed with a solution interval of $\sim$1~min. To get high-precision solutions, we subtracted the structural phase errors of the main phase-referencing calibrator and combined the data from all the sub-bands. To gain accurate phase interpolations between scans, we smoothed the fringe rate solution using a median filter with a width of half hour. Seventh, the instrumental bandpass shapes were corrected. Eighth, both the amplitude and phase self-calibrations were performed for J1334$+$3044. The output solutions were applied to both VLASS J122755.50$+$334526.9 and \target{}. Finally, a phase self-calibration was performed on the data of the faint calibrator VLASS J122755.50$+$334526.9 and its solutions were transferred to \target{}. All the related \textsc{aips} tasks were called via the \textsc{parseltongue} interface \citep{Kettenis2006} and integrated in a single script. 

The deconvolution was performed in \textsc{difmap} \citep{Shepherd1994}. The calibrator J1334$+$3044 shows a core-jet structure with a total flux density of $125 \pm 6$~mJy at 5.0~GHz. The VLASS J122755.50$+$334526.9 shows a slightly resolved jet structure with a total flux density of $20 \pm 1$~mJy at 5.0~GHz.

\subsection{The HSA data at 1.4 GHz}
\label{subsec:HSAobs}
The imaging results of \target{} from the HSA observations at 1.4~GHz on 2005 May 1 were published by \citet{Wrobel2006}. There were also follow-up HSA observations at 1.4~GHz carried out on 2008 May 4. To get independent imaging results, we reduced the latter HSA data. The observations had a recording data rate of 512 Mbps (8~MHz filters, 2~bit quantisation, 16 sub-bands in dual polarisation). \target{} was observed for $\sim$8~h in the phase-referencing mode with a cycle time of $\sim$5~min. The participating stations were St. Croix (\texttt{SC}), Hancock (\texttt{HN}), Fort Davis (\texttt{FD}), Los Alamos (\texttt{LA}), Pie Town (\texttt{PT}), Kitt Peak (\texttt{KP}), Owens Valley (\texttt{OV}), Brewster (\texttt{BR}), Mauna Kea (\texttt{MK}), Green Bank (\texttt{GB}), phased-up VLA (\texttt{Y27}), Arecibo (\texttt{AR}) and \texttt{EF}. 

The experiment was designed with the same phase-referencing strategy as described by \citet{Wrobel2006}. The correlation was done by the VLBA correlator in two passes. We reduced the data following the VLBA continuum data calibration method recommended by the \textsc{aips} cookbook \footnote{\url{http://www.aips.nrao.edu/cook.html}} in Appendix~A.  Both the HSA and EVN calibration strategy are consistent with each other. For unknown reasons, the station \texttt{Y27} had very noisy amplitude calibration data. To avoid potential undesirable effect on calibration, these input calibration data were not used and the amplitude calibration of \texttt{Y27} were achieved by the amplitude self-calibration. The normalization of the amplitude solutions was done without \texttt{Y27}. The off-source visibility data (about 1~min) at the scan beginning of \texttt{Y27} and \texttt{AR} were also flagged out. We noticed that there were significant residual phase errors ($\sim$100$\degr$) on the baselines to \texttt{AR} in the dirty map. Because of this issue, the data of \texttt{AR} were excluded in the final imaging results. The HSA observations used a different calibrator (J1220$+$3431) from that in the EVN observations, while its position is also consistent with the one in ICRF3 (cf. Table~\ref{tab:pos4cal}). The faint target \target{} was imaged without self-calibration.

\subsection{The Jansky VLA data at 12--18 GHz}
\label{subsec:VLAobs}
\target{} was also observed by \citet{Saikia2018} with the Jansky VLA (project code 16B--189) in A configuration at 12-18 GHz on 2016 December 15. During the broad-band observations, \target{} was observed for 4~min. The calibrator 3C~286 was observed as the primary flux density calibrator \citep{Perley2017}. The phase-referencing calibrator was J1215$+$3448 and had a position error of $10$~mas with respect to the ICRF3. In the image published by \citet{Saikia2018}, it displays a faint two-component structure. To study the two components in detail, we revisited the data. The data reduction was performed using the Common Astronomy Software Applications package \citep[\textsc{casa},][]{McMullin2007}. With the VLA pipeline\footnote{\url{https://science.nrao.edu/facilities/vla/data-processing/pipeline}}, the data were calibrated and processed. 

\subsection{The ALMA data at 237 GHz}
\label{subsec:ALMAobs}
We retrieved data from the ALMA science archive (project code: 2017.1.00572.S). \target{} was observed in Band 6 on three epochs: 2018 March 22 and August 28, and 2019 January 23. The observations were taken with the ALMA 12m array in two configurations providing  nominal angular resolutions of 0$\farcs$8 (extended configuration) and 1$\farcs$9 (compact configuration). Quasars J1221$+$2813 and J1229$+$0203 were used for phase-referencing and amplitude/bandpass calibration purposes, respectively. The details of the observational setup are given in Table~\ref{tab:ALMAobs}. The data corresponding to each of the configurations were calibrated separately by the ALMA pipeline using \textsc{casa} versions 5.1.1-5 (extended-configuration) and 5.4.0-68 (compact-configuration). The calibrated data from both configurations were concatenated. For each of the configurations, as well as for the combined data, line-free channels were used to create continuum images. The resulting total aggregate bandwidth is 6.9~GHz, centred around 237.1~GHz. The rms noise level of the final continuum images ranges from 0.03 to 0.05~mJy~beam$^{-1}$.

\begin{table*}
\caption{Summary of the ALMA experiments of \target{} in 2018--2019 and the calibrator information.}
\label{tab:ALMAobs}
\begin{threeparttable}
\begin{tabular}{ccccccc}
\hline
Date        & Configuration          & Min-Max       & Amplitude/Bandpass\tnote{a} 
                                                                                 & Phase\tnote{b} & TOS\tnote{c} & Number of antennas\\     
            &                        & baseline (m)  & J1229$+$0203              & J1221$+$2813   & (min) & \\ 
\hline     
2018 Mar 22 & extended configuration & 15--784      &  4.8 Jy;  $\alpha$=-0.83   & 373$\pm$3~mJy  & 18 & 46 \\
2018 Aug 28 & extended configuration & 15--782      & 6.0 Jy; $\alpha$=-0.75     & 433$\pm$5~mJy  & 18 & 44 \\ 
2019 Jan 23 & compact configuration  & 15--314      & 3.6 Jy; $\alpha$=-0.91     & 253$\pm$2~mJy  & 7  & 48 \\ 
\hline
\end{tabular}
\begin{tablenotes}
\item[a]Assumed flux and spectral index $\alpha$ at 230.3 GHz.\\
\item[b]Derived flux at 230.3 GHz.\\
\item[c]Time on the science source \target{}.
\end{tablenotes}
\end{threeparttable}
\end{table*}

\begin{table*}
\caption{List of the existing high-accuracy radio and optical coordinates of the AGN in NGC~4395.  }
\label{tab:pos4ngc}
\begin{tabular}{ccccccc}
\hline
Method  & RA                                    & $\sigma_{\rm ra}$ 
                                                         & Dec.                                    &  $\sigma_{\rm dec}$  
                                                                                                            & Reference \\     
        &   (J2000)                             & (mas)  & (J2000)                                & (mas)   & \\
\hline     
HSA 
        & 12$^{\rm h}$25$^{\rm m}$48$\fs$8774   &   6    &  $+$33$\degr$32$\arcmin$48$\farcs$715  & 6       & \citet{Wrobel2006}\\
\textit{Gaia} DR3
        & 12$^{\rm h}$25$^{\rm m}$48$\fs$85994  &  1.4   &  $+$33$\degr$32$\arcmin$48$\farcs$7110 & 1.4     & \citet{Gaia2022} \\
Pan-STARRS1 
        & 12$^{\rm h}$25$^{\rm m}$48$\fs$86002  &  13.4   &  $+$33$\degr$32$\arcmin$48$\farcs$7063 & 4.9    & \citet{Chambers2016} \\
\hline
\end{tabular}
\end{table*}
\section{Multi-frequency imaging results}
\label{sec:results}

\begin{figure*}
\centering
\includegraphics[width=\textwidth]{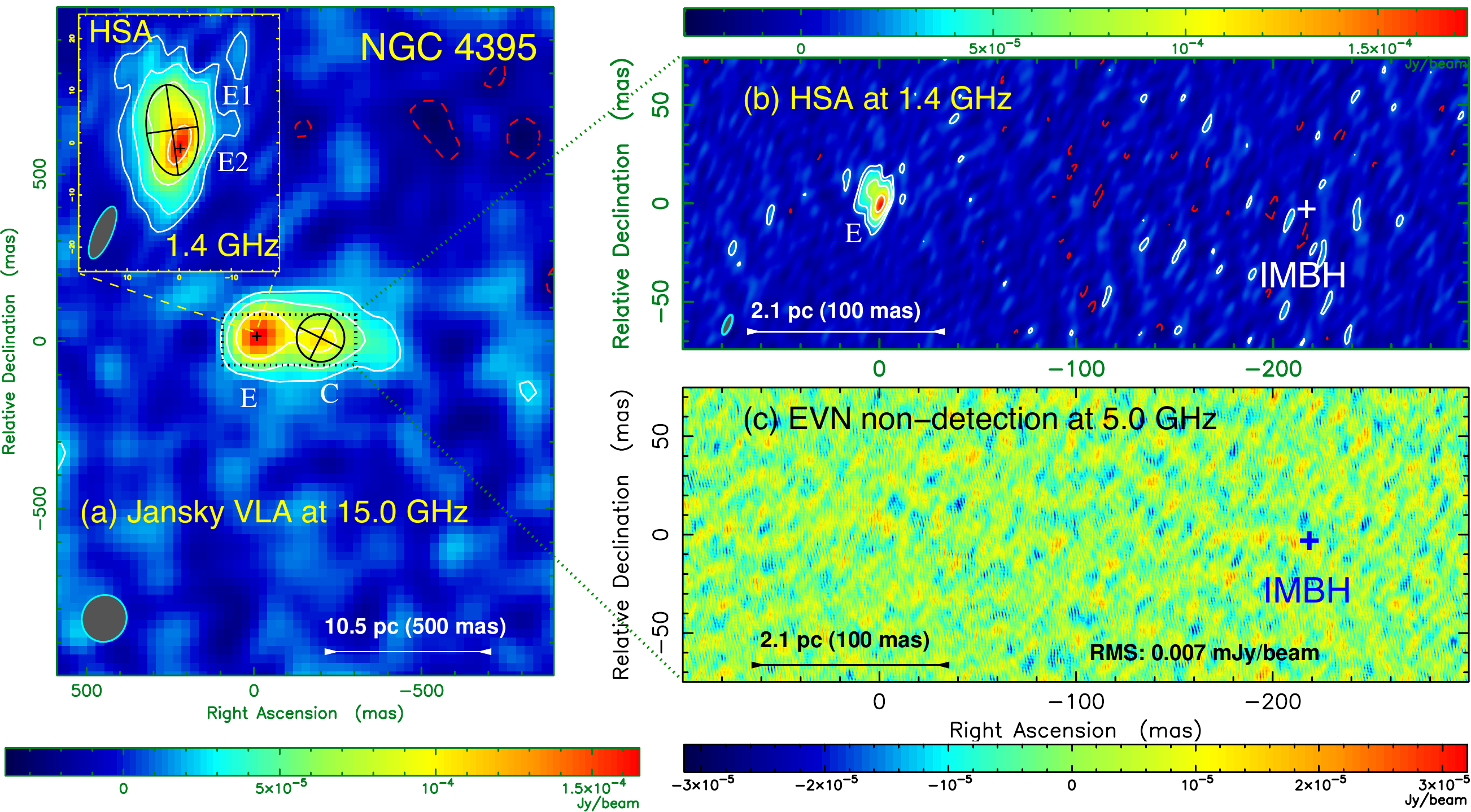}  \\
\caption{
Total intensity images of the nearby low-mass AGN \target{}. North is up and east is left.  All the images were made with natural weighting. In each image, the origin (0, 0) represents the position from the HSA observations of 2005 (cf. Table~\ref{tab:pos4ngc}). The contours give the levels 2.5$\sigma$ $\times$ (-1, 1, 2, 4, ...). The ellipse in the bottom-left corner plots the shape of the beam FWHM. In the VLBI images, the white and blue pluses mark the optical \textit{Gaia} position (cf. Table~\ref{tab:pos4ngc}). \textbf{(a)} The 15-GHz total intensity image made via fitting the VLA visibility data \citep{Saikia2018} to a point-source model (E) plus a circular Gaussian model (C). The first contours are 0.023~mJy\,beam$^{-1}$ (2.5$\sigma$). The peak brightness is 0.165~mJy\,beam$^{-1}$. The FWHM of the synthesised beam is $141 \times 133$~mas. The inset shows the HSA intensity image made via fitting the visibility data to an elliptical Gaussian model (E1) and a point source model (E2). In the images, these models are also plotted as the black pluses, the ellipse and the circle to show their positions and sizes (cf. Table~\ref{tab:modelfit}). \textbf{(b)} The HSA 1.4-GHz \textsc{clean} image. The fist contours are 0.017~mJy\,beam$^{-1}$ (2.5$\sigma$). The peak brightness is 0.173~mJy\,beam$^{-1}$. The beam FWHM is 10.8~$\times$~3.6~mas at position angle $-22\fdg8$.  \textbf{(c)} The EVN 5.0-GHz dirty image. The image has a sensitivity of 0.007~mJy~beam$^{-1}$ and a resolution of 3.13~$\times$~0.95~mas. }
\label{fig:vla_vlbi}
\end{figure*}

\begin{table*}
\caption{Summary of the model fitting results of the HSA and VLA visibility data in \textsc{difmap}. For the relative RA and Dec., we take the early HSA position \citep{Wrobel2006} as the reference point. In the last column, we use the letters EG for the elliptical Gaussian model, PS for the points source model, and CG for the circular Gaussian model.   }
\label{tab:modelfit}
\begin{tabular}{ccccccccccc}
\hline
Label  & Array  & Freq.  & Delta RA      & Delta~Dec.      & $S_{\rm int}$   & $\theta_{\rm maj}$  
                                                                                              & $\theta_{\rm min}$ 
                                                                                                               & $\theta_{\rm pa}$  
                                                                                                                              & $\alpha$     & Model \\
       &       & (GHz)  & (mas)          &  (mas)         & (mJy)            & (mas)          & (mas)          & ($\degr$)    &  ~           & type  \\
\hline
E1     & HSA   & 1.41   & $+1.34\pm0.21$ & $+2.47\pm0.53$ & $0.652\pm0.035$  & $17.5\pm0.9$   & $9.9\pm1.7$    & $7.7\pm3.7$  & ~            & EG    \\
E2     & HSA   & 1.41   & $-0.23\pm0.25$ & $-1.05\pm0.47$ & $0.069\pm0.009$  &  ~             &    ~           &   ~          & ~            & PS    \\
E      & HSA   & 1.41   & $+0.87\pm0.14$ & $+1.35\pm0.36$ & $0.699\pm0.033$  & $15.8\pm0.7$   & $8.9\pm1.3$    & $7.7\pm3.2$  &              & EG    \\ 
E      & VLA   & 15.00  & $-8.52\pm3.60$ & $15.59\pm3.60$ & $0.158\pm0.011$  &  ~             &    ~           &   ~          & $-1.6\pm0.5$ & PS    \\  
C      & VLA   & 15.00  & $-198.76\pm9.20$ & $9.87\pm9.20$ & $0.213\pm0.022$ & $143.0\pm17.0$ & $143.0\pm17.0$ &   ~          & $+0.6\pm0.7$ & CG    \\ 
\hline
\end{tabular} \\
\end{table*}

Table~\ref{tab:pos4ngc} lists the accurate positions of \target{} reported by the previous HSA phase-referencing observations at 1.4~GHz \citep{Wrobel2006}, the optical \textit{Gaia} astrometry \citep{Gaia2021_EDR3, Gaia2022}, and the Panoramic Survey Telescope and Rapid Response System \citep[Pan-STARRS1,][]{Chambers2016}. The systematic positional error (1.4~mas), i.e. the excess noise reported by the \textit{Gaia} data release 3 (DR3) due to various factors (source structure et al.), is included in the error budget. The Pan-STARRS1 position uncertainty contains a systematic uncertainty that came from a comparison of the \textit{Gaia} and Pan-STARRS1 catalogues \citep{Chambers2016}.

\begin{figure*}
\centering
\includegraphics[width=0.3\textwidth]{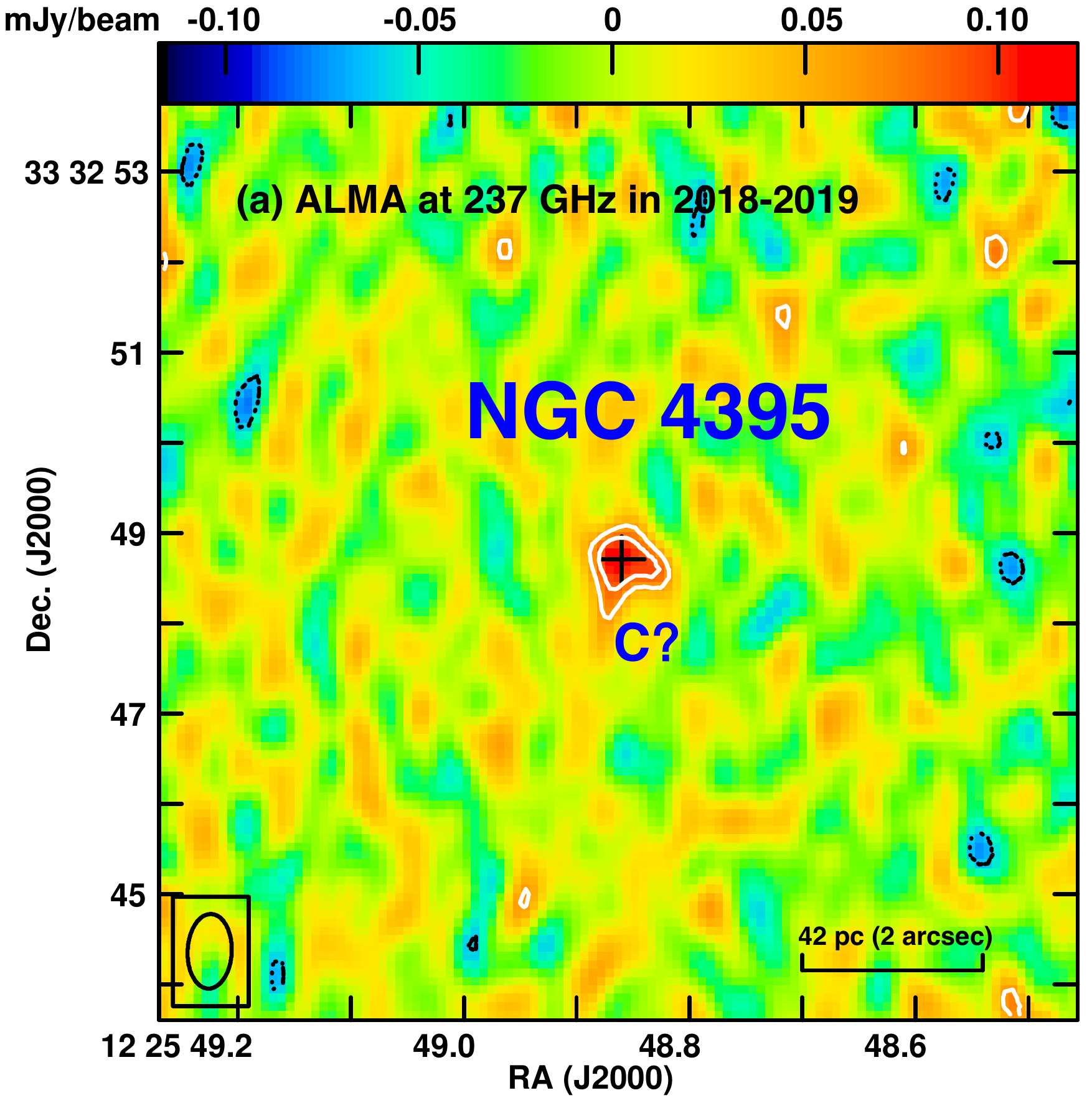}  
\includegraphics[width=0.31\textwidth]{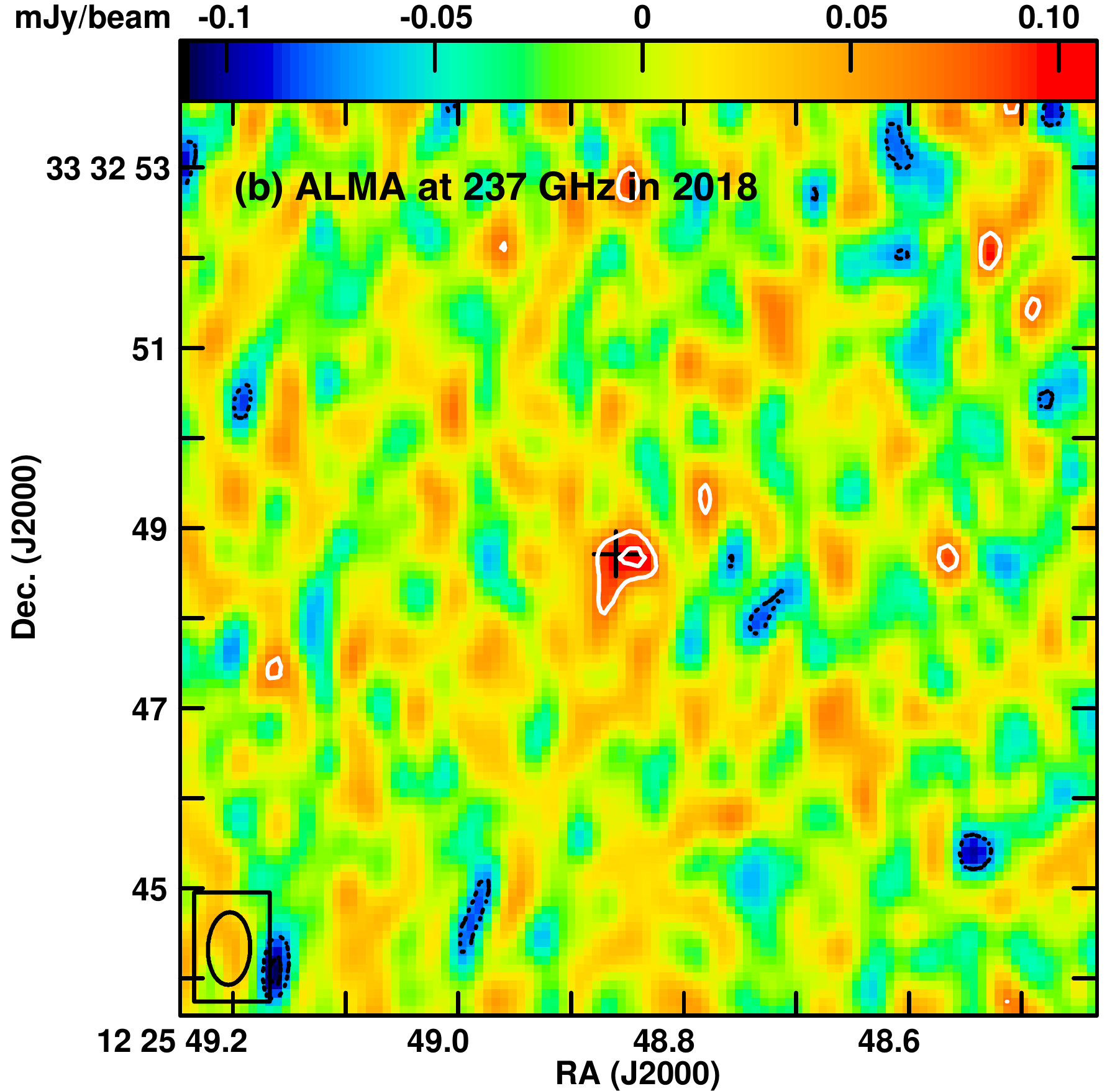}  
\includegraphics[width=0.295\textwidth]{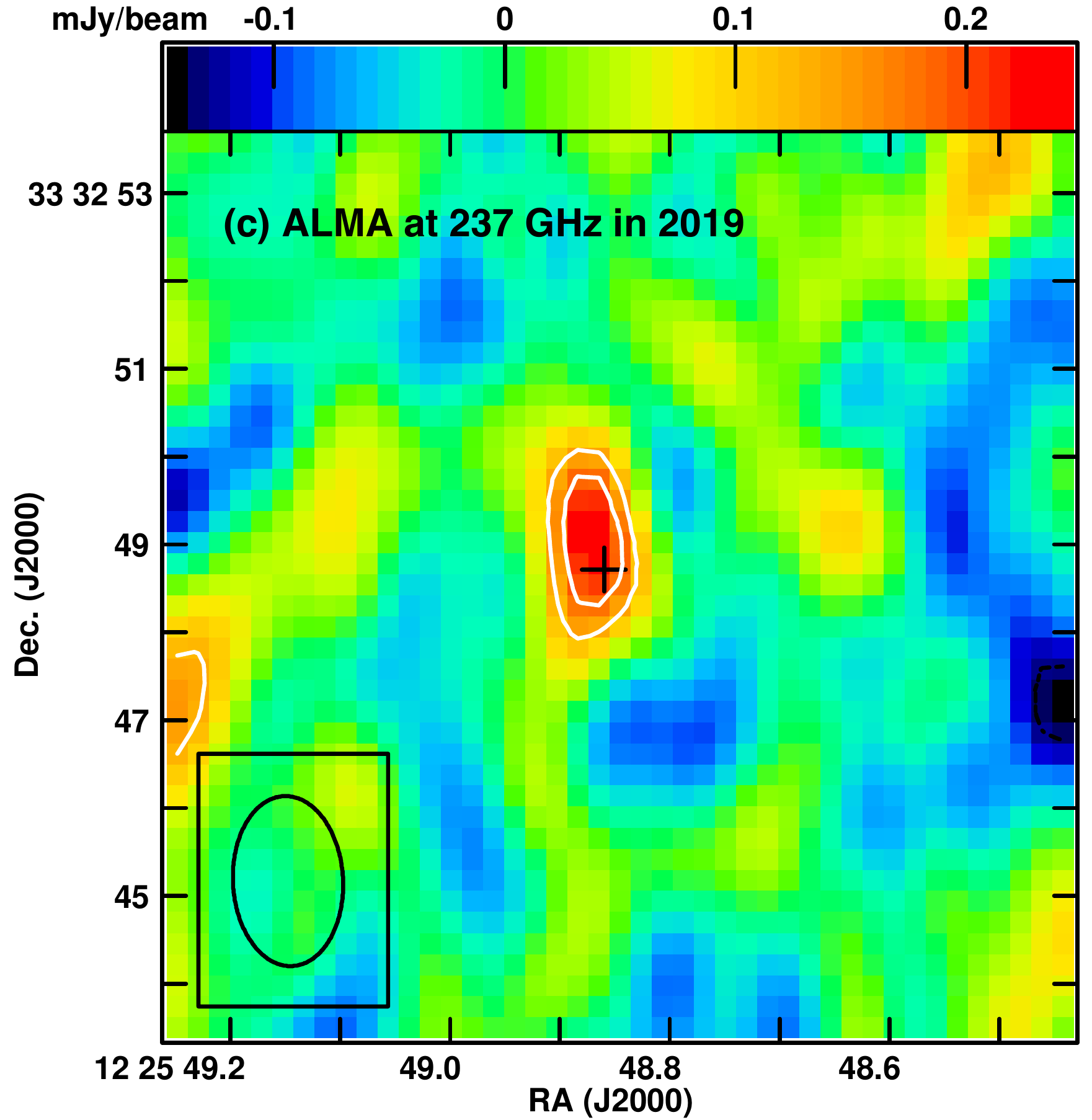}  \\
\caption{
The continuum images of the \target{} nucleus observed by the ALMA at 237.1~GHz. The black plus gives the optical \textit{Gaia} position. (a) The intensity map made by us via combining all the visibility data from 2018--2019 and using a robust parameter of 0.5 in \textsc{casa}. The contours start from 0.061~mJy\,beam$^{-1}$ (2.5$\sigma$) and increase by factors $-$1, 1, 1.4. The peak brightness is 0.112~mJy\,beam$^{-1}$. The beam FWHM is $0.83 \times 0.49$~arcsec at position angle $-2\fdg5$. (b) The archival continuum map from the high-resolution observations on 2018 March 22 and August 28. The contours start from 0.071~mJy\,beam$^{-1}$ (2.5$\sigma$) and increase by factors $-$1.4, $-$1, 1, 1.4. The peak brightness is 0.107~mJy\,beam$^{-1}$. The beam FWHM is $0.81 \times 0.47$~arcsec at position angle $-2\fdg1$. (c) The archival continuum map from the low-resolution observations on 2019 January 23. The contours start from 0.128~mJy\,beam$^{-1}$ (2.5$\sigma$) and increase by factors $-$1, 1, 1.4. The peak brightness is 0.231~mJy\,beam$^{-1}$. The beam FWHM is $1.94 \times 1.25$~arcsec at position angle $+2\fdg9$. }
\label{fig:alma}
\end{figure*}

\begin{figure}
\centering
\includegraphics[width=0.95\columnwidth]{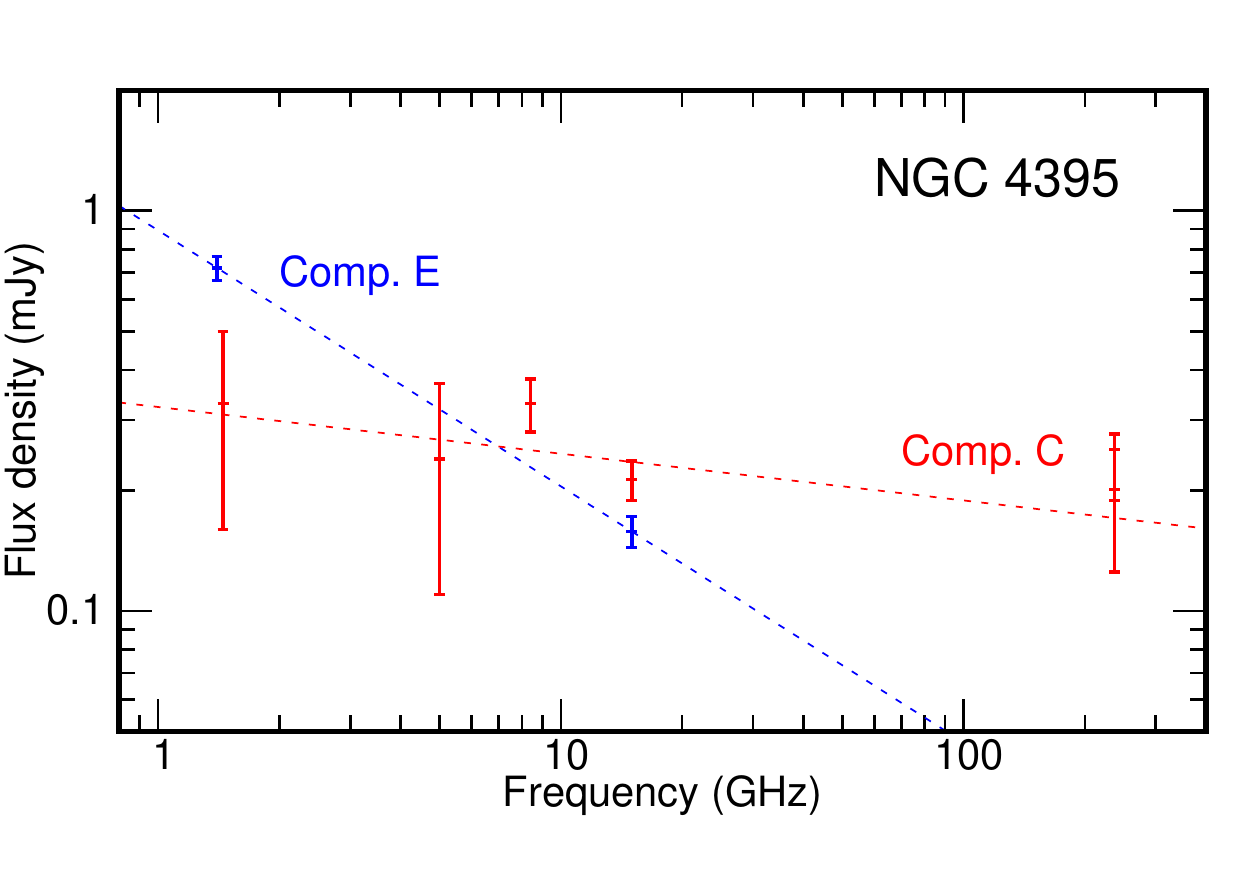}  \\
\caption{
The non-simultaneous radio spectra of the components E and C in the radio nucleus of \target{}. The error bars are at the level 1$\sigma$ and include a 5 per cent systematic error. The spectral indices and the 1-$\sigma$ uncertainties are $\alpha=-0.64\pm0.05$ for the component E (2 blue data points) and $\alpha=-0.12\pm0.08$ for the component C (6 red data points). The data points of the component C at 1.4--8.4 GHz are estimated based on the total flux densities of the radio nucleus reported in literature and the radio spectrum of the component E (cf. section \ref{subsec:nature} for more details). }
\label{fig:spectra}
\end{figure}

Fig.~\ref{fig:vla_vlbi} displays the total intensity images of \target{} observed with the VLA at 12--18~GHz, the HSA at 1.4~GHz and the EVN at 5~GHz. These images have a scale of 0.021~pc\,mas$^{-1}$ and an origin at the position observed by \citep{Wrobel2006} with the HSA. The radio nucleus has been resolved into two relatively discrete features, labelled as E and C, in the VLA image. The component E can be divided into a very extended sub-component E1 and a faint compact sub-component E2 in the high-resolution HSA image. Fig.~\ref{fig:alma} shows the ALMA detection of \target{} at 237 GHz. To characterise these components, we also fit the data to point-source and Gaussian models. During the fitting process, we started to fit each relatively discrete image feature to a simple point-source model. If the point-source model gave some $\la5\sigma$ noise peaks in the residual map, we would fit the feature to the more complex circular or elliptical Gaussian model. In case of the broad-band VLA observations, we also included the spectral index in the models. The fitting results including the formal uncertainties are summarised in Table~\ref{tab:modelfit}. The formal uncertainties are estimated via normalising the reduced $\chi^2$ to unity. The flux densities for each component are also plotted in Fig.~\ref{fig:spectra}. 

With the VLA observations at 12--18 GHz, \citet{Saikia2018} revealed an elongated structure in the radio nucleus of \target{}. Our independent data reduction and imaging analysis have fully confirmed those imaging results. The component C is located at the \textit{Gaia} position, shows a diffuse structure with a size of $143\pm17$~mas ($3.0\pm0.3$~pc) and has an in-band spectral index $\alpha=+0.6\pm0.7$. The component E has an in-band spectral index of $\alpha=-1.6\pm0.5$. The estimates of the in-band spectral indices have a large uncertainty because of the very limited observing bandwidth and the faintness of the two components. Only the component E is clearly detected in the HSA image. 

The wide-field VLBI images are displayed in the right panels of Fig.~\ref{fig:vla_vlbi}. The \textit{Gaia} position \citep{Gaia2021_EDR3} is marked as a white plus and has a separation of 218~mas (4.6~pc) with respect to the HSA position \citep{Wrobel2006}. The HSA image of 2008 in Fig.~\ref{fig:vla_vlbi} has an image quality significantly better the HSA image of 2005 \citep{Wrobel2006}. Owing to the long baselines to \texttt{EF}, the beam area synthesised with natural weighting has become 2.6 times smaller. Because of the two-times higher observing bandwidth, the image sensitivity has been improved by a factor of 1.4. The outflow-like feature E is clearly detected with a signal to noise ratio (SNR) of about 25. The component E is significantly resolved with a size of $17.5 \pm 0.9$~mas ($0.35 \pm 0.02$~pc). Due to the very high image resolution, its peak brightness has decreased to 0.173~\mjyb{} and it is hard to directly see the diffuse emission around the peak feature in the \textsc{clean} map. However, this is easily seen in the residual map after the peak feature is subtracted. With an elliptical Gaussian model (E1) and a point source model (E2), we recover a total flux density of $0.72 \pm 0.05$~mJy, where the error is the quadratic sum of a 5 per cent systematic error and the formal error. The two models are also displayed in the inset of Fig.~\ref{fig:vla_vlbi}a. The sub-component E1 is a very faint feature and only contributes 10 per cent of total flux density of the component E. Adding the faint sub-component E1 allows us to accurately reveal the peak feature of the component E.  Compared to the \textsc{clean} image, this image is made with the model fitting and recovers slightly more diffuse emission on the short baselines and plainly displays the large-scale diffuse emission. To determine the centroid of the entire structure E, we also fit the visibility data to a single elliptical Gaussian model. This is not a very accurate fitting because it gives some significant noise peaks in the residual map and a slightly low total flux density. All these results are consistent with the HSA imaging results reported previously by \citet{Wrobel2006} and further confirms that it is a non-thermal component with an average brightness temperature of $T_{\rm b}=(2.3\pm0.4) \times10^6$~K at 1.4~GHz. \citet{Wrobel2006} also briefly mentioned a 5-$\sigma$ candidate feature near ($<$50~mas) the \textit{Gaia} position, while its existence cannot be confirmed in the 1.4 times more sensitive image.

The component E is not detected at a level of SNR $>$5 in the 5-GHz EVN image. This is because of the steep spectrum and the absence of a compact feature. Assuming no significant time variability, we can derive a spectral index of $\alpha=-0.64\pm0.05$ for the component E in Fig.~\ref{fig:spectra}. This is in agreement with the early report $\alpha=-0.60\pm0.08$ observed by \citet{Ho2001} between 1.4 and 5 GHz with their arcsec-resolution images. We have also searched the region covering the \textit{Gaia} position but did not detect the component C. The two new VLBI images have reached a quite high sensitivity, $\sim$0.007~\mjyb{}. This gives us a 5$\sigma$ upper limit of 0.035~\mjyb{} ($T_{\rm b}= 5.9\times10^5$~K) for the peak brightness at 5~GHz of any compact feature.

In the ALMA archival continuum images at 237~GHz, Fig.~\ref{fig:alma}, \target{} is detected at SNR = 3.8 on 2018 March 22 and SNR = 4.5 on 2019 January 23.  Its total flux densities are $0.20 \pm 0.08$ and $0.19 \pm 0.06$ mJy. To localise the feature precisely, we also combined their visibility data and made a new image, shown in Fig.~\ref{fig:alma}a. The black plus denotes the optical \textit{Gaia} position. Within the 1-sigma error ellipse ($\sigma_{\rm ra} = 0.062$~mas, $\sigma_{\rm dec} = 0.086$~mas), its position is consistent with the \textit{Gaia} position. The average flux density is $0.18 \pm 0.06$~mJy. It is a slightly resolved feature with a size of ($0.89 \pm 0.21$) $\times$ ($0.72 \pm 0.14$)~mas. Because of the very limited image quality (SNR = 4.6), the size estimate may not be reliable.

\section{Discussion}
\label{sec:discussion}

\subsection{Location of the accreting IMBH}
\label{subsec:location}
The IMBH displaying significant optical variability \citep[e.g.][]{Woo2019} is inferred to be located at the \textit{Gaia} position (cf. Table~\ref{tab:pos4ngc}). The \textit{Gaia} photometry shows that \target{} has a mean magnitude of $m_{\rm g}=16.7$. Generally, the multi-epoch full-sky \textit{Gaia} astrometry has reached a position uncertainty of 0.05~mas for compact objects with $m_{\rm g} = 17.0$~mag in the catalogue of the DR3 \citep{Gaia2021_EDR3, Gaia2022}. The sub-arcsec-resolution optical observations of \target{} show that it has a compact nucleus with a radius $<$100~mas \citep[e.g.][]{Matthews1999, Brum2019}. The nucleus is also significantly brighter than any other objects in the g-band optical image of Pan-STARRS1 with a field size of $20\times20$ arcsec. Thus, its optical centroid can be unambiguously and accurately located by the long-term \textit{Gaia} astrometry. In Table~\ref{tab:pos4ngc}, the independent astrometry from the optical Pan-STARRS1 survey fully supports the \textit{Gaia} astrometry. The positional offset between the two optical positions is very small (5~mas) and within the 1$\sigma$ error ellipse. 

The accreting IMBH cannot be hosted by the component E because of its extremely large angular offset (218~mas) to the \textit{Gaia} position. The HSA and EVN images have a precision of $\sim$1~mas with respect to the optical \textit{Gaia} frame. The two phase-referencing calibrators that are used to locate \target{} in the VLBI observations also have the precise \textit{Gaia} positions available. The separations between the \text{Gaia} and ICRF3 positions are small, 1.2~mas for J1220$+$3431 and 0.2~mas for J1220$+$3343. Moreover, the \textit{Gaia} astrometry of compact AGNs generally has no significant systematic errors of $>$10~mas with respect to the radio cores of VLBI catalogues \citep[e.g.][]{Petrov2019, Yang2020IRAS, Titov2022}.

\subsection{Nature of the components E and C}
\label{subsec:nature}
\target{} is a Seyfert galaxy. In the image from the optical integral field unit (IFU) observations, the ratio [O\textsc{iii}]$\lambda$5007/H$\beta$ is in the range 5.5--11.0 and has a peak at the optical centroid. This supports the entire nuclear region hosting the components C and E being ionised by the central accreting IMBH instead of by star formation \citep{Brum2019}. Moreover, the total star-forming rate is very low, $0.03$ M$_{\sun}$\,yr$^{-1}$ \citep[][]{Smirnova2020}.

The EVN non-detection of the steep-spectrum component E further strengthens the case that it is a very diffuse feature and has no compact jet base. The Jansky VLA monitoring observations at 8.4~GHz show that it has no significant variability on timescales of days \citep{King2013}. The arcsec-scale compact radio nucleus was first observed by \citet{Sramek1992} and \citet{Moran1999} with the A-configuration VLA at 5.0~GHz on 1982 February 8 and had a flux density of $0.56\pm0.12$~mJy. The second-epoch VLASS survey reported that it had flux densities $0.82\pm0.24$~mJy at 2--4 GHz on 2020 September 12 \citep{Lacy2020}. Assuming a stable radio spectrum with a spectral index of $\alpha=-0.60\pm0.08$ \citep{Ho2001}, the recent VLASS detection supports there being no large flux density variability over about four decades. Due to the relatively stable flux density and the non-circular structure, the component E cannot be naturally identified as a young supernova or a supernova remnant.  The possibility of it being a compact starburst has been excluded because of its high brightness temperature $\sim$10$^6$~K and no evidence of starbursts in the nuclear region \citep{Brum2019}. Using the high-resolution \textit{Hubble Space Telescope}, \citet{Woo2019} identified a biconical [O\textsc{iii}] outflow along the east-west direction in \target{}.  The position of the component E is consistent with the outer edge of the outflow. Furthermore, there exists some high ionization and high density molecular gas with a size $\la$1~arcsec \citep[see Fig. 5 \& 13, ][]{Brum2019} in the eastern region. In view of these observational findings, the non-thermal and steep-spectrum feature E probably represents sub-pc-scale shocks. These shocks may provide additional energy to heat the surrounding gas and strengthen the H$_2$ and [Fe\textsc{ii}] emission lines imaged by \citet{Brum2019} with the optical IFU observations. Shocks are also found in NGC~404 \citep{Nyland2017}. The shocks in NGC~404 have a size up to 20 pc at 12--18 GHz, show an optically thin radio spectrum and likely originate from an outflow.

These shocks in \target{} might result from a continuous outflow from the IMBH or from fading ejecta launched by an episode of IMBH activity. The very low brightness temperature $(2.3\pm0.4) \times 10^6$~K indicates no significant Doppler beaming effect \citep[e.g.][]{Readhead1994}. The non-detection of the significant offset between the two HSA observations supports that the component E has a very low transverse speed, $<$0.01~$c$, where $c$ is the light speed.

The sub-mJy ALMA feature is likely associated with the component C instead of the component E. Its centroid is consistent with the optical \textit{Gaia} position (cf. Section~\ref{sec:results}). Furthermore, its flux densities are tentatively consistent with the low-frequency spectrum of the component C. Fig.~\ref{fig:spectra} shows the non-simultaneous radio spectra of the components E and C. Using the radio spectrum of the component E and the total flux densities observed by the \textit{e}-MERLIN at 1.4 GHz \citep{Baldi2021} and the VLA at 5 and 8.4 GHz \citep{Sramek1992, Moran1999, King2013}, we also derived the flux densities of the component C at frequencies $\leq$8.4~GHz. The indirect measurements are also reliable because these images have sub-arcsec resolutions and the entire radio nucleus is not significantly resolved. The three data points are also plotted in Fig.~\ref{fig:spectra}. Because the component E has a steep spectrum with $\alpha=-0.64\pm0.05$, the data points at 237~GHz are quite consistent with the extrapolation of the spectra of the component C. There is no evidence for a strong absorption at the lower observing frequencies. The broad-band radio spectrum can be described by a single power law function. The best-fitting model for the non-simultaneous spectrum between 1.4 and 237 GHz is $S_{\nu} = (0.32\pm0.08) \nu^{-0.12\pm0.08}$. Because of the large uncertainty (10--50 percent) of these flux density measurements, the spectral index estimate is only sensitive to large ($\ga$30 per cent) flux density variability. We caution that the estimate also has a certain bias because of different image resolutions and possible non-uniform absorption at very low frequencies. Moreover, the component C might have a more complex radio spectrum. Future VLA and ALMA observations, at matched angular resolutions, could add the more data points between 15.0 and 237.1 GHz and thus test the assumption of the single power law spectrum.  

The central pc-scale component C possibly originates from thermal free-free emission instead of non-thermal synchrotron emission.  Firstly, its average brightness temperatures are very low: $T_{\rm b}=56\pm9$~K in the 15~GHz VLA map and $\leq5.9\times10^5$~K in the 1.4-GHz HSA and 5-GHz EVN maps. Secondly, its radio morphology is very extended, $143\pm17.0$~mas at 15~GHz. There is no sub-pc-scale compact component detected in the VLBI maps. Finally, its broad-band radio spectrum tends to be relatively flat ($\alpha = -0.12\pm0.08$). A flat radio spectrum can also come from a partially optically thick synchrotron jet base \citep[cf. a recent review by][]{Blandford2019}. However, the jet base will have a quite high brightness temperature of $>10^{6}$~K and show a conical structure. 

If the diffuse component C is a thermal emission region, it may result from a clumpy torus \citep[e.g.][]{Netzer2015, GamezRosas2022, Isbell2022}. The diffuse feature has a radius of $\sim$1.4~pc at 15 GHz and its centroid coincides with the IMBH. As a clumpy torus including a polar outflow, it is slightly larger than that observed in the archetypal type 2 galaxy NGC~1068 \citep[e.g.][]{Gallimore2004, GamezRosas2022}, while $\sim$200 times smaller than the first-known thick torus directly imaged in Cygnus A \citep{Carilli2019}. \citet{Combes2019} studied the molecular tori of seven nearby Seyfert galaxies and reported the radii in the range 6--27 pc. Compared to these Seyfert galaxies, \target{} is expected to have a relatively small torus \citep[cf. a review by][]{Netzer2015} because of its very low $M_{\rm bh}$ \citep[e.g.][]{Woo2019} and bolometric luminosity \citep[e.g.][]{Brum2019}. The observed radius of $\sim$1.4~pc is consistent with the expected outer edge of the torus, which is naturally considered as the gravitational sphere of influence of the IMBH ($R_{\rm BH,sph}$). By adopting  $M_{\rm bh} = 10^{4-5} M_\odot$ and velocity dispersion $\sigma_\star$ = 18 km~s$^{-1}$ \citep{Woo2019}, the estimated $R_{\rm BH,sph}$ is $\sim$1.0~pc according to Equation 3 in \citet{Netzer2015}. Near infrared IFU observations by \citet{Brum2019} reveal a slightly extended H$_2$ emission line region in the nucleus of \target{}. The molecular H$_2$ region also roughly covers the thermal radio emission region. Because of the limited image sensitivity and the near face-on disc geometry, there is no evidence of a recognisable disc structure in the radio maps. 

\subsection{Implications}
\label{subsec:implications}

A sub-mJy mas-scale compact radio core is frequently seen at $\la$5~GHz in AGNs \citep[e.g.][]{Deller2014, HerreraRuiz2017, Fischer2021, Yang2021}. Its existence represents a continuous and partially optically thick jet base associated with the accreting massive black hole \citep[e.g.][]{Blandford2019}. If there is a compact jet base formed by the central accreting IMBH of \target{}, the HSA and EVN non-detections at the \textit{Gaia} position allow us to set the 5$\sigma$ upper limits on its luminosity, $L_{\rm R} = \nu L_{\nu} = 1.3\times10^{33}$~erg\,s$^{-1}$ at 1.4~GHz and $L_{\rm R} = 4.7\times10^{33}$~erg\,s$^{-1}$ at 5.0~GHz. For nearby AGNs \citep[e.g.][]{Giroletti2009, Baldi2021, Fischer2021}, radio non-detections can generally reach a luminosity upper limit of $\sim10^{35}$~erg\,s$^{-1}$. Thus, the two luminosity limits for \target{} are quite stringent. VLBI non-detections of the radio cores are also reported in low-mass AGNs, e.g. Henize~2--10 \citep{Reines2012} and NGC~404 \citep[][]{Paragi2014}. These VLBI non-detections indicate that the accreting IMBHs have very weak jet activity or stay in a radio quiescent state \citep[e.g.][]{Greene2006RQ}. 

\target{} shows a strong variability in the full X-ray range \citep[e.g.][]{Kammoun2019} and has an absorption-corrected, time-averaged 2--10~keV luminosity, $L_{\rm X}=8.8\times10^{39}$~erg\,s$^{-1}$ \citep{Moran2005}. However, \citet{King2013} reported that there is no significant correlation between the radio and X-ray light curves observed simultaneously by the VLA and \textit{Swift} for about two months. Our VLBI non-detection of the jet base is fully in agreement with the very weak correlation because the VLA observations trace the pc-scale extended emission instead of the sub-pc-scale compact jet base near the IMBH. 

The fundamental plane of BH activity is a statistical relation derived from an incomplete sample of hard-state stellar-mass BHs in X-ray binaries \citep[e.g.][]{Fender2004} and supermassive BHs in AGNs. It represents a correlation among X-ray luminosity, radio luminosity, and BH mass, and shows a very large scatter ($\la$1~dex). This plane potentially supports the existence of a common physical mechanism acting across the mass scales \citep[e.g.][]{Merloni2003, Falcke2004, Saikia2018}. \target{} has been frequently used as a precious data point to fill the mass gap of the plane. However, the recent multi-resolution VLA and VLBA observations of 25 nearby AGNs by \citet{Fischer2021} suggest that the radio luminosity supporting the fundamental plane mainly result from the radio emission integrated on the VLA instead of VLBA scales. The VLBI non-detection of \target{} at 5~GHz allows us to add an extremely low radio luminosity point to the plots of the fundamental plane relation. Because of the relatively large radio luminosity offset ($\ge$0.8 dex) from the fundamental plane relation of \citep{Merloni2003} at 5~GHz, the data point of \target{} in the mass gap tends to support the recent VLBI results reported by \citep{Fischer2021}. The radio luminosity of VLBI-scale compact AGN cores at 5 GHz may not follow the prediction of the fundamental plane relation. 

The component C has a radio luminosity of $L_{\rm R} = (8.6 \pm 1.0) \times 10^{34}$~erg\,s$^{-1}$ at 15~GHz. \target{} is a very low-luminosity AGN \citep{Moran1999}. Its bolometric luminosity $L_{\rm bol}$ is between $1.9 \times 10^{40}$~erg\,s$^{-1}$ and $4.9 \times 10^{41}$~erg\,s$^{-1}$ \citep{Brum2019}. If the thermal nature of the pc-scale component C in such a low-luminosity AGN is confirmed by future radio observations, the large population of the low-luminosity AGNs \citep[cf. a review by][]{Ho2008} would allow us to to find additional similar objects in deep sky surveys. 

\section{Conclusions}
\label{sec:conclusion}

To answer whether there is a sub-pc-scale continuous jet tracing the accreting IMBH in \target{}, we carried out deep EVN observations at 5.0~GHz, reduced HSA data at 1.4~GHz and Jansky VLA data at 12--18~GHz, and examined  ALMA archival continuum images at 237~GHz. We found that the previously known outflow-like feature E displays additional diffuse structure in the 1.4-GHz HSA image of 2008, shows a steep spectrum between 1.4 and 15.0~GHz, and has no compact substructure detected in the 5.0-GHz EVN image. Together with the very large angular offset (about 220 mas) from the mas-accuracy optical \textit{Gaia} position, we suggest that the eastern non-thermal feature E traces nuclear shocks probably due to prior IMBH ejection activity. In the VLA and ALMA images, we also identified a 100-mas-scale feature C surrounding the accreting IMBH. Because of its relatively flat spectrum and VLBI non-detections at 1.4 and 5~GHz, we tentatively interpreted the central pc-scale feature C as a thermal emission region originating from a clumpy torus. Moreover, the deep VLBI images show no evidence for radio emission from a jet base at the \textit{Gaia} position, taken to mark the location of the accreting IMBH. Assuming no strong radio variability, the VLBI non-detection at 5~GHz indicates a very weak jet base at the IMBH, $\leq4.7\times10^{33}$~erg\,s$^{-1}$, and no direct evidence of a sub-pc-scale disc-jet coupling above the luminosity limit. 

The upcoming high-resolution IFU spectroscopy\footnote{\url{https://www.stsci.edu/jwst/science-execution/program-information.html?id=2016}} with the \textit{James Webb Space Telescope} would significantly advance our understanding on the inner structure of the infrared nucleus of \target{} \citep{Seth2021}. In the future, the next-generation VLA (ngVLA\footnote{\url{https://ngvla.nrao.edu/}}) observations of \target{} would reach an image sensitivity of 0.1~$\mu$Jy\,beam$^{-1}$ (2-h on-source time, 8-GHz bandwidth) and a resolution of $\sim$1 mas (maximum baseline length 8860 km with a long-baseline array) at 4--12~GHz, possibly revealing an IMBH jet base.  

\section*{Acknowledgements}
LCH was supported by the National Science Foundation of China (11721303, 11991052, 12011540375) and the China Manned Space Project (CMS-CSST-2021-A04).
LF gratefully acknowledges the support of NSFC (12173037), Cyrus Chun Ying Tang Foundations and the Strategic Priority Research Program of Chinese Academy of Sciences, Grant No. XDB 41010105.
The EVN is a joint facility of independent European, African, Asian, and North American radio astronomy institutes. Scientific results from data presented in this publication are derived from the following EVN project code(s): EY039. 
The National Radio Astronomy Observatory is a facility of the National Science Foundation operated under cooperative agreement by Associated Universities, Inc.
e-VLBI research infrastructure in Europe is supported by the European Union’s Seventh Framework Programme (FP7/2007-2013) under grant agreement number RI-261525 NEXPReS.
\textit{e}-MERLIN is a National Facility operated by the University of Manchester at Jodrell Bank Observatory on behalf of STFC.
This work has made use of data from the European Space Agency (ESA) mission {\it Gaia} (\url{https://www.cosmos.esa.int/gaia}), processed by the {\it Gaia} Data Processing and Analysis Consortium (DPAC, \url{https://www.cosmos.esa.int/web/gaia/dpac/consortium}). Funding for the DPAC has been provided by national institutions, in particular the institutions participating in the {\it Gaia} Multilateral Agreement.
This research has made use of the NASA/IPAC Extragalactic Database (NED), which is operated by the Jet Propulsion Laboratory, California Institute of Technology, under contract with the National Aeronautics and Space Administration.
This research has made use of NASA’s Astrophysics Data System Bibliographic Services. 
This research has made use of the VizieR catalogue access tool, CDS, Strasbourg, France (DOI : 10.26093/cds/vizier). The original description  of the VizieR service was published in 2000, A\&AS 143, 23.
This research has made use of the CIRADA cutout service at URL cutouts.cirada.ca, operated by the Canadian Initiative for Radio Astronomy Data Analysis (CIRADA). CIRADA is funded by a grant from the Canada Foundation for Innovation 2017 Innovation Fund (Project 35999), as well as by the Provinces of Ontario, British Columbia, Alberta, Manitoba and Quebec, in collaboration with the National Research Council of Canada, the US National Radio Astronomy Observatory and Australia’s Commonwealth Scientific and Industrial Research Organisation.
Basic research in radio astronomy at the U.S. Naval Research Laboratory is supported by 6.1 Base Funding. 
This paper makes use of the following ALMA data: ADS/JAO.ALMA\#2017.1.00572.S. ALMA is a partnership of ESO (representing its member states), NSF (USA) and NINS (Japan), together with NRC (Canada), MOST and ASIAA (Taiwan), and KASI (Republic of Korea), in cooperation with the Republic of Chile. The Joint ALMA Observatory is operated by ESO, AUI/NRAO and NAOJ. 
\section*{Data Availability}
The correlation data underlying this article are available in the EVN Data Archive (\url{http://www.jive.nl/select-experiment}) and the NRAO Science Data Archive. The calibrated visibility data underlying this article will be shared on reasonable request to the corresponding author.




\bibliographystyle{mnras}
\bibliography{NGC4395_2021} 







\bsp	
\label{lastpage}
\end{document}